\documentclass[aps,amsmath,amssymb,preprint,groupedaddress,superscriptaddress,floatfix]{revtex4-1}
\usepackage[T2A]{fontenc}
\usepackage[utf8]{inputenc}
\usepackage[english]{babel}

\usepackage{graphicx}
\usepackage{nicefrac}
\usepackage{amsmath,amsfonts,amssymb,amsthm,mathtools}
\mathtoolsset{showonlyrefs=false}
\usepackage{epsf}
\usepackage{bm}
\usepackage{braket}
\usepackage{dcolumn}
\usepackage{multirow}
\usepackage{booktabs}
\usepackage{hyperref}
\usepackage{float}
\usepackage{bbold}
\usepackage{subcaption}

\newcommand{\etal}{\textit{et al.}}

\begin{document}
\title{One-electron energy spectra of heavy highly charged quasimolecules}


\author{Artem~A.~Kotov}
\affiliation{Department of Physics, St. Petersburg State University, 199034 St. Petersburg, Russia}

\author{Dmitry~A.~Glazov}
\affiliation{Department of Physics, St. Petersburg State University, 199034 St. Petersburg, Russia}

\author{Vladimir~M.~Shabaev}
\affiliation{Department of Physics, St. Petersburg State University, 199034 St. Petersburg, Russia}

\author{G\"unter~Plunien}
\affiliation{Institut f\"ur Theoretische Physik, \\ Technische Universit\"at Dresden, D-01062 Dresden, Germany}

\begin{abstract}
The generalized dual-kinetic-balance approach for axially symmetric systems is employed to solve the two-center Dirac problem. The spectra of one-electron homonuclear quasimolecules are calculated and compared with the previous calculations. The analysis of the monopole approximation with two different choices of the origin is performed. Special attention is paid to the lead and xenon dimers, Pb$^{82+}$--Pb$^{82+}$--e$^{-}$ and Xe$^{54+}$--Xe$^{54+}$--e$^{-}$, where the energies of the ground and several excited $\sigma$-states are presented in the wide range of internuclear distances. The developed method provides the quasicomplete finite basis set and allows for construction of the perturbation theory, including within the bound-state QED. 
\end{abstract}
\maketitle

\section{Introduction}
Due to the critical phenomena of the bound-state quantum electrodynamics, such as spontaneous electron-positron pair production, quasimolecular systems emerging in ion-ion or ion-atom collisions attract much interest \cite{Gerstein69, Pieper69,Zeldovich72,Rafelski78,Greiner1985,Maltsev2019,Popov2020,Voskresensky2021}.
While collisions of highly charged ions with neutral atoms are presently available for experimental investigations, in particular, at the GSI Helmholtz Center for Heavy Ion Research \cite{Verma2006_245,Verma2006_75,Hagmann2011}, the upcoming experiments at the GSI/FAIR~\cite{Gumberidze2009}, NICA~\cite{TerAkopian2015}, and HIAF~\cite{Ma2017} facilities will allow observation of the heavy ion-ion (up to U$^{92+}$--U$^{92+}$) collisions.
The relativistic dynamics of the heavy-ion collisions has been investigated for decades by various methods, see, e.g., Refs.~\cite{Soff79,Becker86,Eichler90,Rumrich93,Ionescu99,Tupitsyn2010,Tupitsyn2012,Maltsev2019,Popov2020,Voskresensky2021} and references therein. Theoretical predictions of the quasimolecular spectra are also in demand for analysis of the experimental data in these collisions. 

Within the Bohr-Oppenheimer approximation, the one-electron problem is reduced to the Dirac equation with Coulomb potential of two nuclei at a fixed internuclear distance $D$. This problem was investigated previously by a number of authors, see, e.g., Refs.~\cite{Muller1973,Rafelski76,Rafelski76_PRL,Lisin1977,Soff79,Lisin1980,Yang91,Parpia95,Deineka1998,Matveev2000,Kullie2001,Ishikawa2008,Tupitsyn2010,Artemyev2010,Ishikawa2012,Tupitsyn2014,Mironova2015,Artemyev2015}. Majority of these works relied on the partial-wave expansion of the two-center potential in the center-of-mass coordinate system. Alternative approaches include, e.g., usage of the Cassini coordinates~\cite{Artemyev2010} and the atomic Dirac-Sturm basis-set expansion~\cite{Tupitsyn2014,Mironova2015}.
We consider the method based on the dual-kinetic-balanced finite-basis-set expansion \cite{Shabaev2004} of the electron wave function for the axially symmetric systems \cite{Rozenbaum2014}. The results for the ground state of uranium dimers with one and two electrons were already presented in Ref.~\cite{Kotov2020}. In this work, we extend the one-electron calculations to the lowest excited $\sigma$-states and present the results for the one-electron dimers, Pb$^{82+}$--Pb$^{82+}$--e$^{-}$ and Xe$^{54+}$--Xe$^{54+}$--e$^{-}$. For the ground state we demonstrate the accuracy of this method for the nuclear charge numbers $Z$ from 1 to 100 at the so-called ``chemical'' distances, $D = 2/Z$ a.u. We also investigate the difference between the two-center values and those obtained within the monopole approximation. 

The relativistic units ($ \hbar = 1 $, $ c = 1 $, $ m_e = 1 $) and the Heaviside charge unit ($ \alpha = e^2/(4\pi), e < 0 $) are used throughout the paper.

\section{Method}
In heavy atomic systems the parameter $ \alpha Z $ ($ \alpha $ is the fine-structure constant and $ Z $ is the nuclear charge), which measures the coupling of electrons with nuclei, is not small. Therefore, the calculations for these systems should be done within the fully relativistic approach, i.e., to all orders in $ \alpha Z $. With this in mind, we start with the Dirac equation for the two-center potential,
\begin{gather}
	\label{eq:dirac_eq}
	\Big[ \vec{\alpha}\cdot\vec{p} + \beta + V(\vec{r}) \Big] \Psi_n(\vec{r}) = E_n \Psi_n(\vec{r}), \\
	V(\vec{r}) = V^{\text{A}}_{\text{nucl}}(|\vec{r}-\vec{R}_1|) + V^{\text{B}}_{\text{nucl}}(|\vec{r}-\vec{R}_2|).
	\label{eq:TC_pot}
\end{gather}
Here $ \vec{r} $ and $ \vec{R}_{1,2} $ are the coordinates of the electron and nuclei, respectively, $ V^{\text{A,B}}_{\text{nucl}}(r) $ is the nuclear potential at the distance $ r $ generated by nucleus with the charge $ Z_{\text{A,B}} $, $ \vec{\alpha} $ and $ \beta $ are the standard $ 4 \times 4 $ Dirac matrices:
\begin{equation}
	\beta = \begin{pmatrix}
		         I & 0 \\
        		 0 & -I
        	\end{pmatrix},
	\qquad\vec{\alpha} = 
    	\begin{pmatrix}
    		0 & \vec{\sigma} \\
    		\vec{\sigma} & 0
    	\end{pmatrix},
\end{equation}
where $ \vec{\sigma} $ is a vector of the Pauli matrices.

In the following we consider the identical nuclei, i.e. $ Z_{\text{A}} = Z_{\text{B}} $, with the Fermi model of the nuclear charge distribution:
\begin{equation}
	V_{\text{nucl}}(r)=-4\pi\alpha Z\,\int\limits_{0}^{\infty}\frac{\rho(r')}{\max(r,r')}r'^2\,dr', \qquad \rho(r) = \frac{\rho_0}{1+\exp{(r-c)/a}},
\end{equation}
where $ \rho_0 $ is the normalization constant, $ a $ is skin thickness constant and $ c $ is the half-density radius, for more details see, e.g., Ref.~\cite{Shabaev1993}.

The solution of Eq.~\eqref{eq:dirac_eq} is obtained within the dual-kinetic-balance (DKB) approach, which allows one to solve the problem of ``spurious'' states.
Originally, this approach was implemented for spherically symmetric systems, like atoms, \cite{Shabaev2004}, using the finite basis set constructed from the B-splines \cite{Johnson1988,Sapirstein1996}.
Later, authors of Ref.~\cite{Rozenbaum2014} generalized it to the case of axially symmetric systems (A-DKB) --- they considered atom in the external homogeneous field. This situation was also considered within this method in Refs.~\cite{Varentsova2017,Volchkova2017,Volchkova2021} to evaluate the higher-order contributions to the Zeeman splitting in highly charged ions. In Ref.~\cite{Kotov2020} we have adapted the A-DKB method to diatomic systems, which also possess axial symmetry.
Below we provide a brief description of the calculation scheme.

The system under consideration is rotationally invariant with respect to the $z$-axis directed along the internuclear vector $ \vec{D} = \vec{R}_2 - \vec{R}_1$. Therefore, the $z$-projection of the total angular momentum with the quantum number $m_J$ is conserved and the electronic wave function can be written as,
\begin{equation}
\label{eq:psi_function}
	\Psi(r, \theta, \varphi) = \frac{1}{r} \begin{pmatrix}
		 G_1(r,\theta) e^{i(m_J - \frac{1}{2})\varphi} \\
		 G_2(r,\theta) e^{i(m_J + \frac{1}{2})\varphi} \\
		iF_1(r,\theta) e^{i(m_J - \frac{1}{2})\varphi} \\
		iF_2(r,\theta) e^{i(m_J + \frac{1}{2})\varphi}
	\end{pmatrix}\,.
\end{equation}
The $(r, \theta)$-components of the wave function are represented using the finite-basis-set expansion:
\begin{equation}
	\Phi(r,\theta) = \frac{1}{r} \begin{pmatrix}
		G_1(r,\theta) \\
		G_2(r,\theta) \\
		F_1(r,\theta) \\
		F_2(r,\theta)
	\end{pmatrix} \cong \sum\limits_{u=1}^{4}\sum\limits_{i_r=1}^{N_r}\sum\limits_{i_{\theta}=1}^{N_{\theta}} C_{i_r,i_{\theta}}^u\Lambda B_{i_r}(r) Q_{i_{\theta}}(\theta) e_u\,,
	\label{eq:bse}
\end{equation}
where
$ \big\{ B_{i_r}(r) \big\}_{i_r=1}^{N_r} $ are B-splines, $ \big\{ Q_{i_{\theta}} \big\}_{i_{\theta}=1}^{N_{\theta}} $ are Legendre polynomials of the argument $ 2\theta/\pi - 1 $, and $ \big\{ e_u \big\}_{u=1}^{4} $ are the standard four-component basis vectors:
\begin{equation}
	e_1 = 
	\begin{pmatrix}
		1 \\
		0 \\
		0 \\
		0
	\end{pmatrix},\quad
	e_2 = 
	\begin{pmatrix}
    	0 \\
    	1 \\
    	0 \\
    	0
    \end{pmatrix},\quad
	e_3 = 
	\begin{pmatrix}
    	0 \\
    	0 \\
    	1 \\
    	0
    \end{pmatrix},\quad
	e_4 = 
	\begin{pmatrix}
    	0 \\
    	0 \\
    	0 \\
    	1
    \end{pmatrix}\,.
\end{equation}
The $ \Lambda $-matrix 
\begin{gather}
	\Lambda = \begin{pmatrix}
                          I & -\frac{1}{2}D_{m_J} \\
		-\frac{1}{2}D_{m_J} & I
	\end{pmatrix}, \\
	D_{m_J} = (\sigma_z\cos\theta + \sigma_x\sin\theta)\left( \frac{\partial}{\partial r} - \frac{1}{r} \right)
\nonumber\\
	+ \frac{1}{r}(\sigma_x\cos\theta - \sigma_z\sin\theta)\frac{\partial}{\partial\theta}
	+ \frac{1}{r\sin\theta}\left( im_J\sigma_y + \frac{1}{2}\sigma_x \right)
	\,,
\end{gather}
imposes the dual-kinetic-balance conditions on the basis set. With the given form of $\Phi$ and the finite basis set one can find the corresponding Hamiltonian matrix $H_{ij}$. The eigenvalues and eigenfunctions are found by diagonalization of $H_{ij}$. As a result, we obtain quasicomplete finite set of wave functions and electronic energies for the two-center Dirac equation. Ground and several lowest excited states are reproduced with high accuracy while the higher-excited states effectively represent the infinite remainder of the spectrum. The negative-energy continuum is also represented by the finite number of the negative energy eigenvalues. This quasicomplete spectrum can be used to construct the Green function, which is needed for the perturbation theory calculations.

\section{Results}
Relativistic calculations of the binding energies of heavy one-electron quasimolecules were presented, in particular, in Refs.~\cite{Parpia95,Kullie2001,Artemyev2010,Tupitsyn2010,Tupitsyn2014,Mironova2015}, see also references therein. Ref.~\cite{Mironova2015} provides nearly the most accurate up-to-date values for the very broad range of $Z$ and taking into account the finite nuclear size. So, we use just these data for comparison, see Table~\ref{t:Mironova_comp}, where the ground-state energies are presented for $Z=1\dots100$ at the so-called ``chemical'' distances, $D = 2/Z$ a.u. We observe that the results are in good agreement, the relative deviation varies from $2\times 10^{-6}$ for hydrogen to $5\times 10^{-5}$ for $Z=100$. This deviation is consistent with our own estimation of the numerical uncertainty, which is evaluated by inspecting the convergence of the results with respect to the size of the basis set. In this calculation up to $ N_r = 320 $ B-splines and $ N_{\theta} = 54 $ Legendre polynomials are used, for heavy nuclei this number of basis functions ensures the uncertainty, which is comparable to or smaller than the uncertainty of the finite nuclear size effect at all internuclear distances from $0$ to $2/Z$ a.u.

\begin{table}
\caption{Comparison of the ground-state energies of one-electron quasimolecular systems with $Z=1\dots100$ at the internuclear distance $D = 2 / Z$ a.u.}
\label{t:Mironova_comp}
\setlength{\tabcolsep}{2mm}
\centering
\begin{tabular}{rr@{}lr@{}l}
\toprule
$ Z $ 
& \multicolumn{2}{c}{This work} 
& \multicolumn{2}{c}{Dirac-Sturm \cite{Mironova2015}} \\
\midrule
  1 & $-$1.&1026433 & $-$1.&102641581032 \\
  2 & $-$4.&4106607 & $-$4.&410654714140 \\
 10 & $-$110.&33722 & $-$110.&3371741499 \\
 20 & $-$442.&23969 & $-$442.&2392996469 \\
 30 & $-$998.&4194 & $-$998.&4214646525 \\
 40 & $-$1783.&5479 & $-$1783.&563450815 \\
 50 & $-$2804.&5304 & $-$2804.&571434254 \\
 60 & $-$4070.&971 & $-$4071.&036267926 \\
 70 & $-$5595.&889 & $-$5595.&926978290 \\
 80 & $-$7397.&003 & $-$7397.&028800116 \\
 90 & $-$9498.&452 & $-$9498.&588788490 \\
 92 & $-$9957.&567 & $-$9957.&775519122 \\
100 & $-$11935.&89 & $-$11936.&41770218 \\
\bottomrule
\end{tabular}
\end{table}

Next, we present the obtained one-electron spectra of the Pb$^{82+}$--Pb$^{82+}$--e$^{-}$ and Xe$^{54+}$--Xe$^{54+}$--e$^{-}$ quasimolecules in the wide range of the internuclear distances from few tens of fermi up to the ``chemical'' distances.
In the present figures only $\sigma$-states ($m_J = \pm \frac{1}{2}$) are shown.
The precise quantum numbers are $ m_J $ and parity, \textit{g} (\textit{gerade}) or \textit{u} (\textit{ungerade}).
In addition, we determine the quantum numbers of the ``merged atom'', i.e. the state of the system with internuclear distance $ D \to 0 $, and put it to the left of molecular term symbol, e.g., the ground state is $ 1s_{1/2} \sigma_g $.

\begin{figure}
    \begin{subfigure}[b]{0.49\textwidth}
        \centering
        \includegraphics[width=\textwidth]{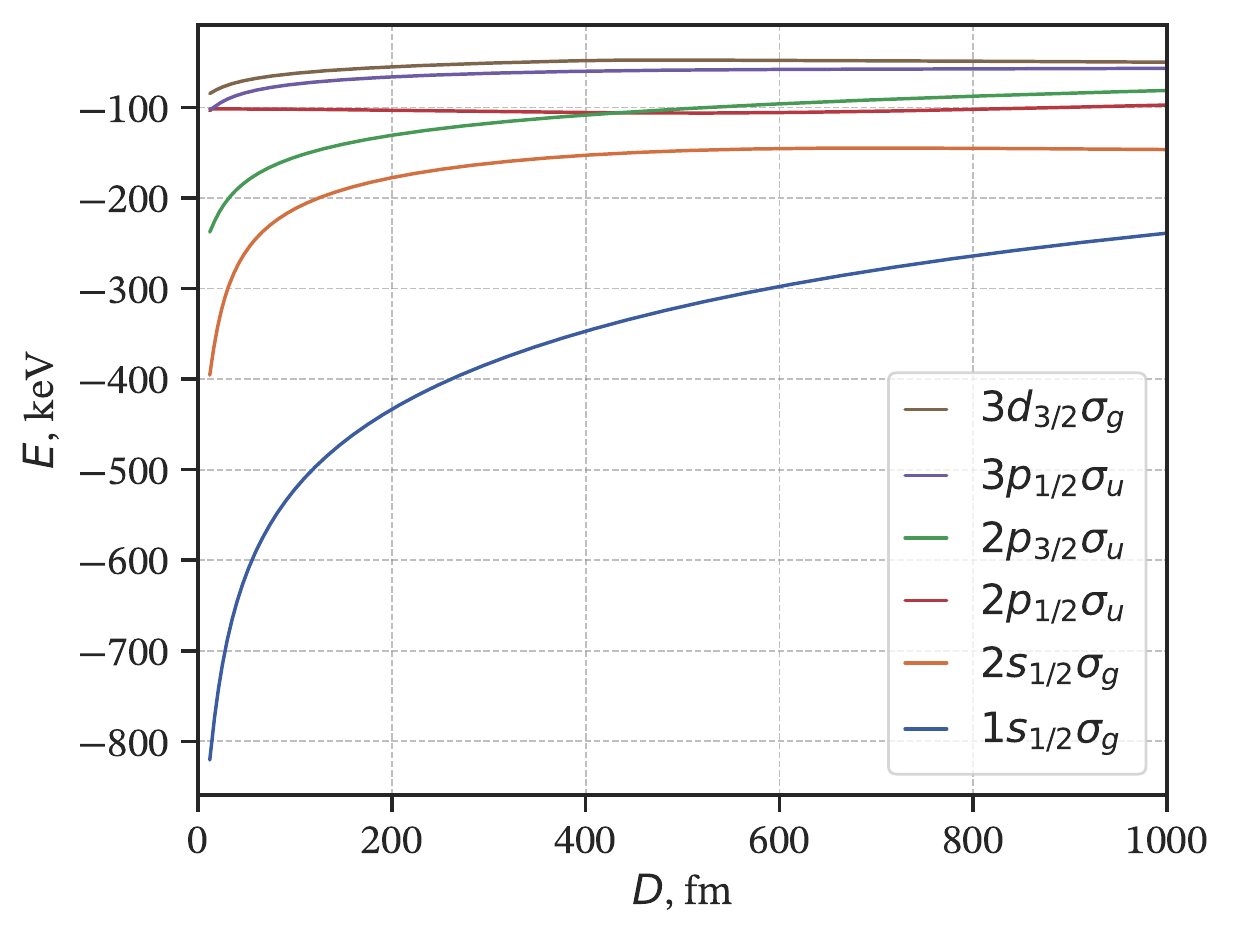}
        \caption{Energies of states with $n=1\ldots 6$.}
        \label{fig:Pb_a}
    \end{subfigure}
    \hfill
    \begin{subfigure}[b]{0.49\textwidth}
        \includegraphics[width=\textwidth]{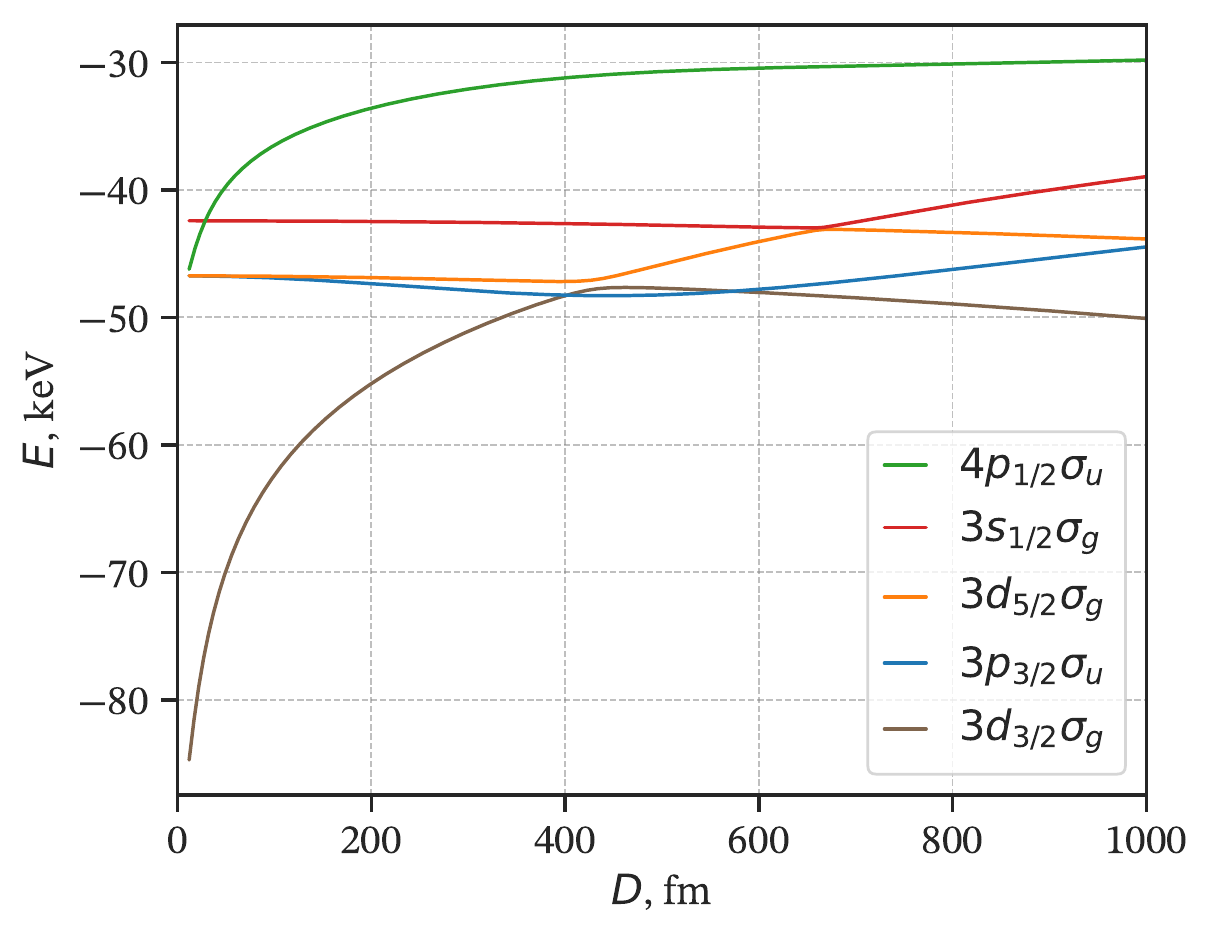}
        \caption{Energies of states with $n=6\ldots 10$.}
        \label{fig:Pb_b}
    \end{subfigure}
    \caption{Electronic terms of the one-electron Pb$^{82+}$--Pb$^{82+}$ quasimolecule.}
    \label{fig:Pb}
\end{figure}

\begin{figure}
    \begin{subfigure}[b]{0.49\textwidth}
        \centering
        \includegraphics[width=\textwidth]{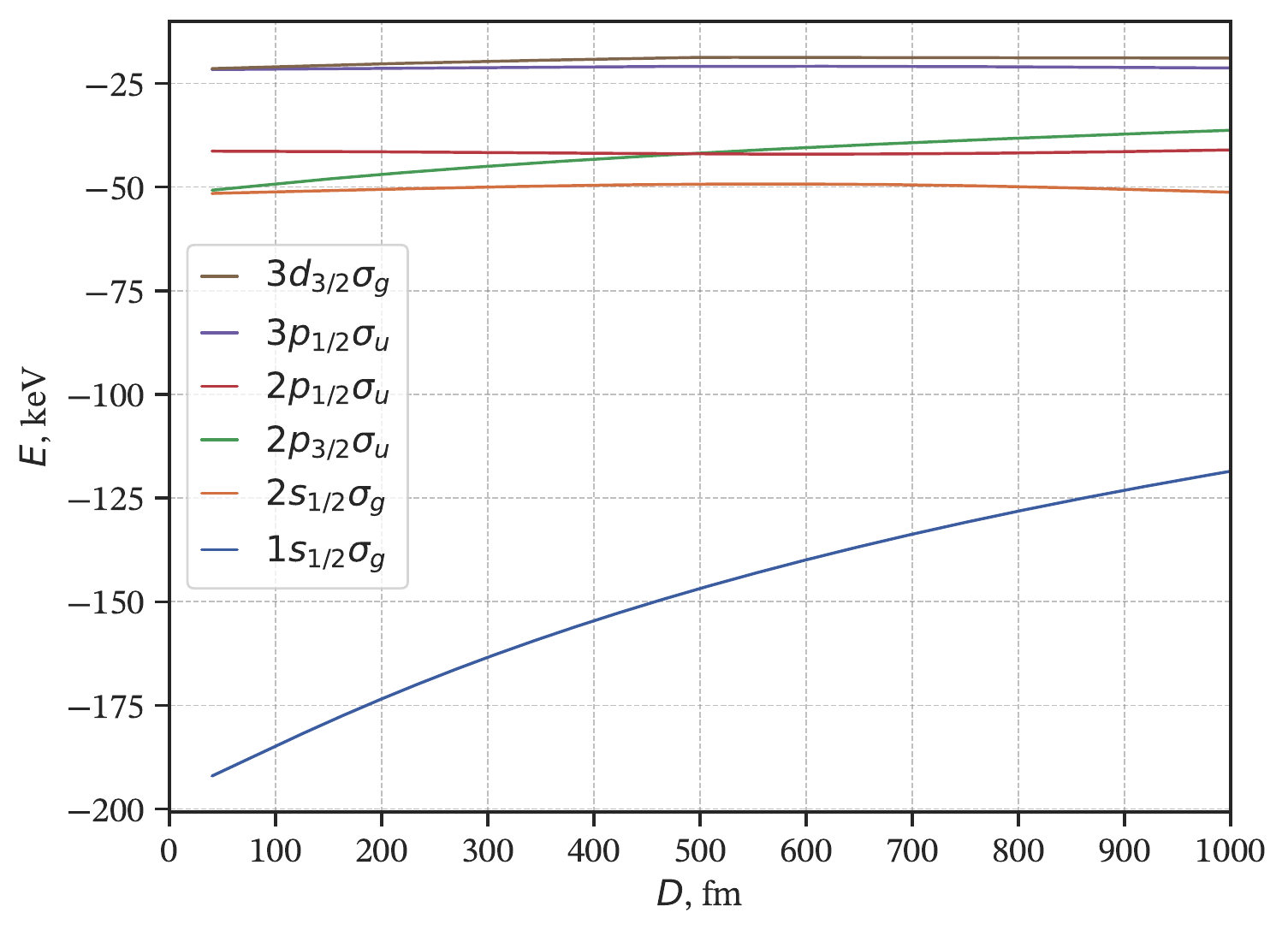}
        \caption{Energies of states with $n = 1 \ldots 6$.}
        \label{fig:Xe_a}
    \end{subfigure}
    \hfill
    \begin{subfigure}[b]{0.49\textwidth}
        \includegraphics[width=\textwidth]{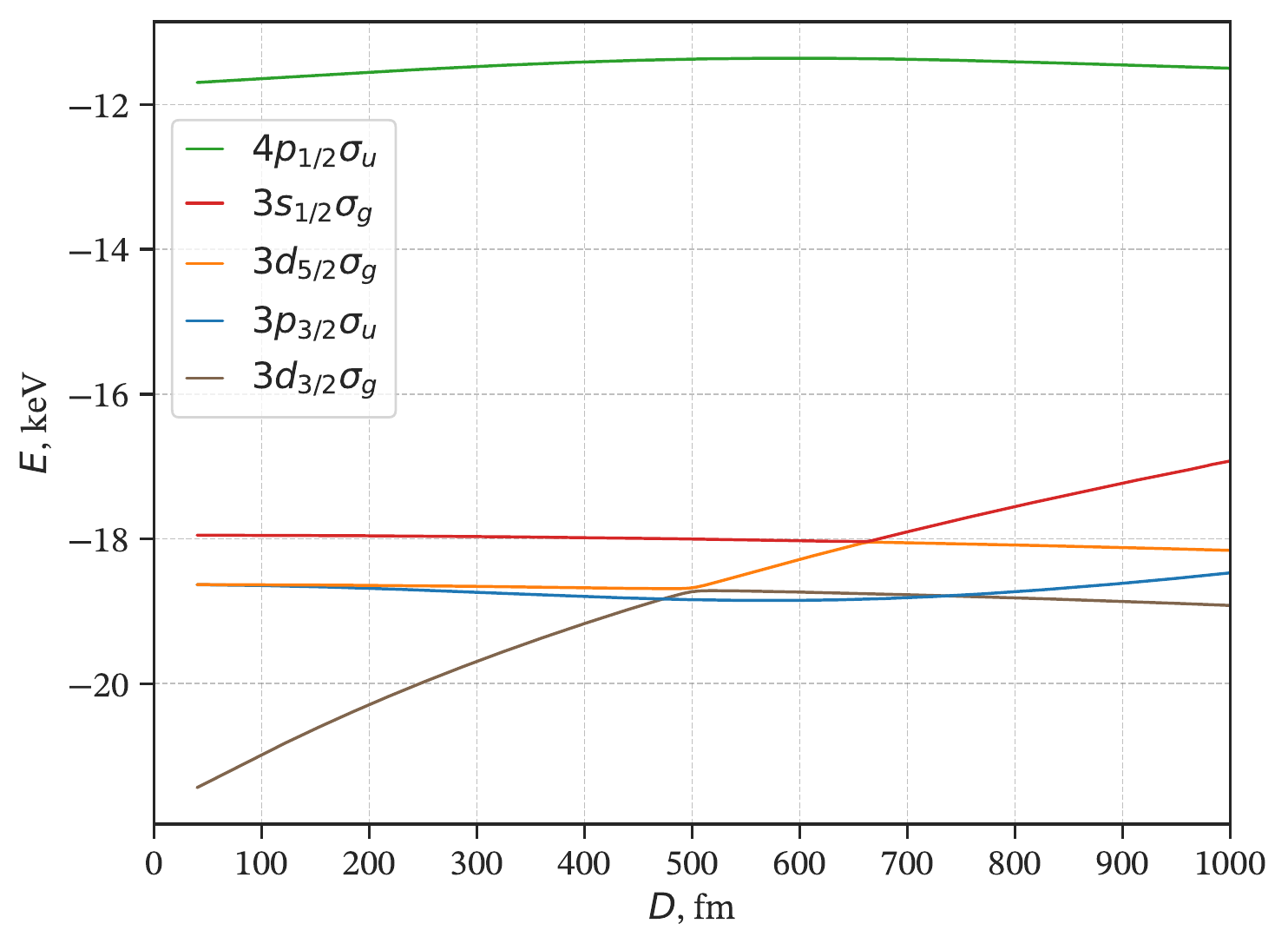}
        \caption{Energies of states with $n = 6 \ldots 10$.}
        \label{fig:Xe_b}
    \end{subfigure}
    \caption{Electronic terms of the one-electron Xe$^{54+}$--Xe$^{54+}$ quasimolecule.}
    \label{fig:Xe}
\end{figure}

In Figure~\ref{fig:Pb}, the energies of the ground ($n=1$) and first 9 ($n=2 \ldots 10$) excited states of Pb$ ^{82+} $--Pb$ ^{82+} $--e$ ^{-} $ system as the functions of the internuclear distance are shown. Here, $n$ has no connection with atomic principal quantum number, it simply enumerates the $\sigma$-states.
%
%
To visually compare the data obtained with the ones by Soff \etal\ we zoom the second plot in Fig.~\ref{fig:Pb} to match the scale of the corresponding figure from Ref.~\cite{Soff79}. Although we cannot compare the numerical results, the plots for all the states under consideration appear to be in very good agreement --- all the states are identified correctly, all the crossings and avoided crossings appear at the same internuclear distances.
\begin{figure}
    \includegraphics[width=0.5\textwidth]{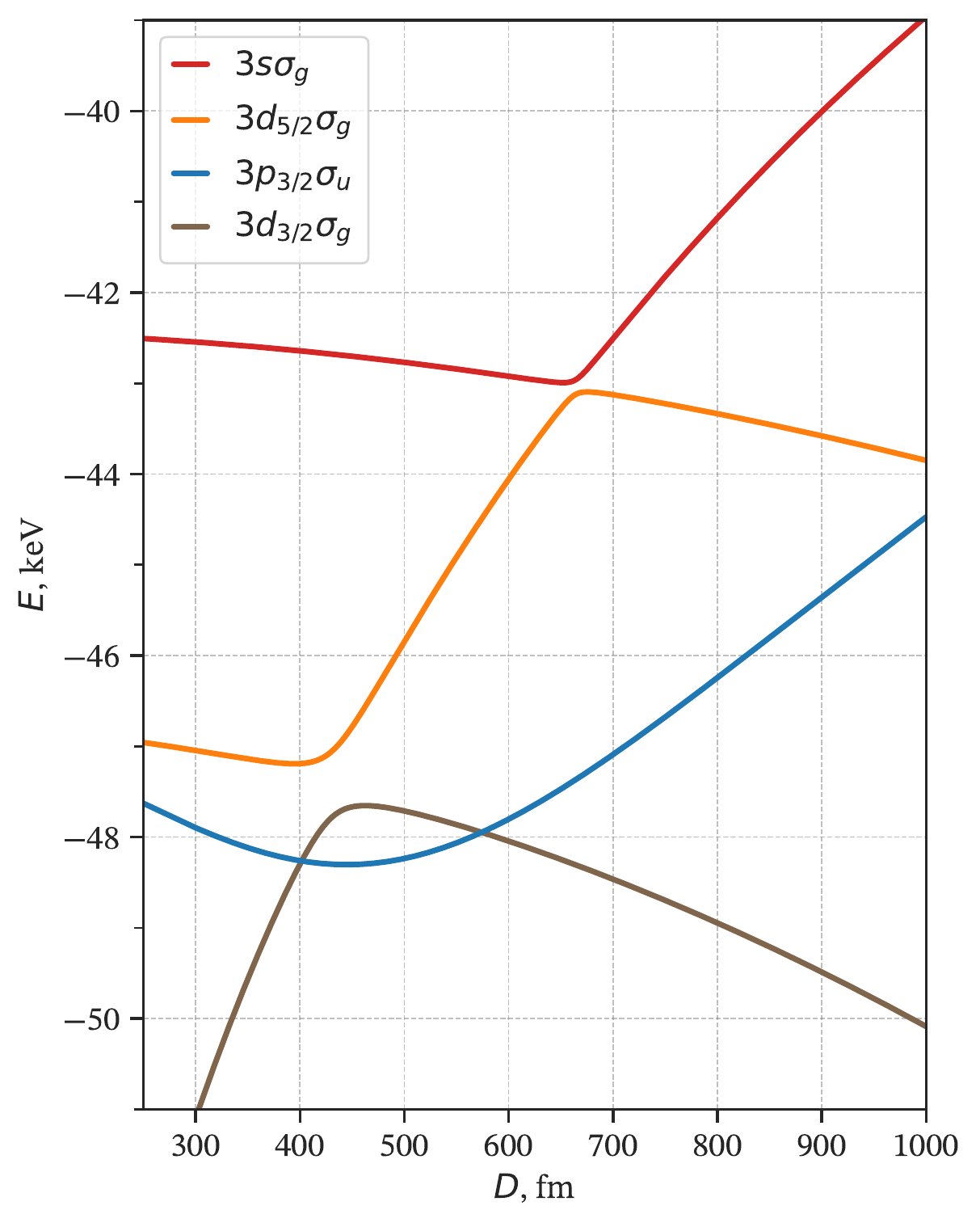}
    \caption{Electronic terms of the one-electron Pb$^{82+}$--Pb$^{82+}$ quasimolecule. Energies of states with $n=6\ldots 9$ (scaled).}
    \label{fig:Pb_Soff}
\end{figure}
The similar results for xenon, i.e., the energies of the ground ($n=1$) and first 9 ($n=2 \ldots 10$) excited states of Xe$ ^{54+} $--Xe$ ^{54+} $--e$ ^{-} $ system are shown in Figure~\ref{fig:Xe}.

Also, in Tables~\ref{t:Pb_TC_MA} and \ref{t:Xe_TC_MA}, we compare the ground-state binding energies obtained within our approach for the two-center (TC) potential with those for the widely used monopole approximation (MA), where only the spherically symmetric part of the two-center potential is considered. Within MA the potential and all the results depend on where to place the origin of the coordinate system (c.s.).
\begin{figure}
    \centering
    \includegraphics[width=0.6\textwidth]{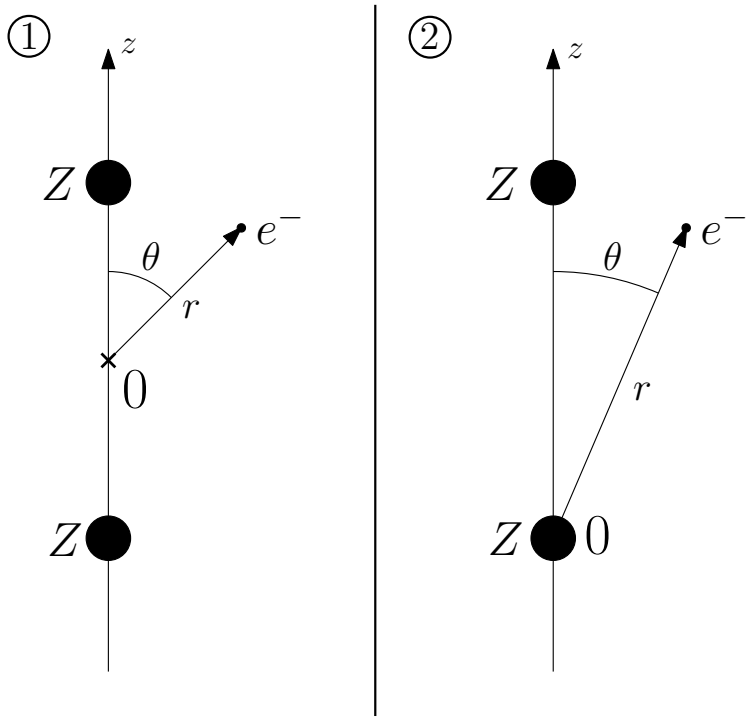}
    \caption{Two different coordinate systems considered. Left: $(1)$ origin is at the center-of-mass of the system, right: $(2)$ origin is at the center of one of the nuclei.}
    \label{fig:diffCS}
\end{figure}
At the same time, for the TC potential the results should be identical within the numerical error bars. We compare two different placements of the c.s. origin: $(1)$ at the center of mass of the nuclei, $(2)$ at the center of one of the nuclei, see Figure~\ref{fig:diffCS}. The agreement between TC(1) and TC(2) within the anticipated numerical uncertainty serves as a non-trivial self-test of the method, since the basis-set expansion~(\ref{eq:bse}) is essentially different for the two cases. In fact, due to the lower symmetry of the second c.s., the uncertainty of the TC(2) values is much larger and completely determines the difference between TC(1) and TC(2). The differences between the TC(1) and MA(1) results are presented in the second-to-last column, they can be interpreted as inaccuracy of the MA. In the last column, the differences between the MA(1) and MA(2) results are given, a kind of ``inherent inconsistency'' of the MA. As one can see from these data, except for the regions where $ \text{MA}(1) - \text{MA}(2) $ is anomaly small due to the sign change, it is comparable to $ \text{TC}(1) - \text{MA}(1) $. This observation can be used to quantify the inaccuracy of the MA for the contributions, which are not yet available for the TC calculations, e.g., the two-photon-exchange and QED corrections~\cite{Kotov2020}.
\begin{table}
\caption{Ground-state binding energy $E_{1\sigma_g}$ (in eV) of the Pb$^{82+}$--Pb$^{82+}$--e$^{-}$ quasimolecule for the two-center potential (TC) and for the monopole-approximation potential (MA), with coordinate system origin at the center of mass of the nuclei $(1)$ and at the one of the nuclear centers $(2)$.}
\label{t:Pb_TC_MA}
\setlength{\tabcolsep}{2mm}
\begin{tabular}{rrrrrrr}
\toprule
$ D $, fm 
& \multicolumn{1}{c}{$ {\text{TC}(1)} $} 
& \multicolumn{1}{c}{$ {\text{TC}(2)} $} 
& \multicolumn{1}{c}{$ {\text{MA}(1)} $} 
& \multicolumn{1}{c}{$ {\text{MA}(2)} $} 
& \multicolumn{1}{c}{$ \text{TC}(1) - \text{MA}(1) $} 
& \multicolumn{1}{c}{$ \text{MA}(1) - \text{MA}(2) $} \\
\midrule
                                40 & $-$646254 & $-$646254 & $-$637032 & $-$598564 &  $-$9222 & $-$38468 \\
                                50 & $-$614504 & $-$614504 & $-$604643 & $-$568188 &  $-$9861 & $-$36455 \\
                                80 & $-$550575 & $-$550575 & $-$539861 & $-$506742 & $-$10714 & $-$33119 \\
                               100 & $-$521373 & $-$521373 & $-$510350 & $-$478423 & $-$11023 & $-$31927 \\
                               200 & $-$433348 & $-$433347 & $-$421146 & $-$392345 & $-$12202 & $-$28801 \\
                               250 & $-$405450 & $-$405450 & $-$392687 & $-$365185 & $-$12763 & $-$27502 \\
                               500 & $-$319773 & $-$319769 & $-$304337 & $-$283510 & $-$15436 & $-$20827 \\
 $ 1/Z\,[\text{a.u.}] \simeq 645 $ & $-$289068 & $-$289067 & $-$272212 & $-$255389 & $-$16856 & $-$16823 \\
                               700 & $-$279462 & $-$279464 & $-$262095 & $-$246756 & $-$17367 & $-$15339 \\
                              1000 & $-$238887 & $-$238873 & $-$218905 & $-$211937 & $-$19982 &  $-$6968 \\
$ 2/Z\,[\text{a.u.}] \simeq 1291 $ & $-$212020 & $-$212003 & $-$189652 & $-$190174 & $-$22368 &    522 \\
\bottomrule
\end{tabular}
\end{table}

\begin{table}
\caption{Ground-state binding energy $E_{1\sigma_g}$ (in eV) of the Xe$^{54+}$--Xe$^{54+}$--e$^{-}$ quasimolecule. The notations are the same as in Table~\ref{t:Pb_TC_MA}.}
\label{t:Xe_TC_MA}
\setlength{\tabcolsep}{2mm}
\begin{tabular}{rrrrrrr}
\toprule
$ D $, fm & $ {\text{TC}(1)} $ & $ {\text{TC}(2)} $ & $ {\text{MA}(1)} $ & $ {\text{MA}(2)} $ & $ \text{TC}(1) - \text{MA}(1) $ & $ \text{MA}(1) - \text{MA}(2) $ \\
\midrule
                               40 & $-$192031 & $-$192031 & $-$191860 & $-$190033 &  $-$171 & $-$1827 \\
                               50 & $-$190845 & $-$190845 & $-$190607 & $-$188314 &  $-$238 & $-$2293 \\
                               80 & $-$187217 & $-$187216 & $-$186775 & $-$183199 &  $-$442 & $-$3576 \\
                              100 & $-$184805 & $-$184805 & $-$184228 & $-$179895 &  $-$577 & $-$4333 \\
                              200 & $-$173425 & $-$173425 & $-$172190 & $-$165031 & $-$1235 & $-$7159 \\
                              250 & $-$168242 & $-$168242 & $-$166695 & $-$158621 & $-$1547 & $-$8074 \\
                              500 & $-$146803 & $-$146802 & $-$143860 & $-$133919 & $-$2943 & $-$9941 \\
                              700 & $-$133710 & $-$133710 & $-$129828 & $-$120118 & $-$3882 & $-$9710 \\
  $1/Z\,[\text{a.u.}] \simeq 980$ & $-$119414 & $-$119413 & $-$114421 & $-$105971 & $-$4993 & $-$8450 \\
                             1000 & $-$118529 & $-$118528 & $-$113464 & $-$105233 & $-$5065 & $-$8231 \\
 $2/Z\,[\text{a.u.}] \simeq 1960$ &  $-$89269 &  $-$89276 &  $-$81433 &  $-$79488 & $-$7836 & $-$1945 \\
\bottomrule
\end{tabular}
\end{table}

\section{Discussion and conclusion}
In this work, the two-center Dirac equation is solved within the dual-kinetic-balance method~\cite{Shabaev2004,Rozenbaum2014}. The energies of the ground and several excited $\sigma$-states in such heavy diatomic systems as Pb$^{82+}$--Pb$^{82+}$--e$^{-}$ and Xe$^{54+}$--Xe$^{54+}$--e$^{-}$ are plotted as a function of the internuclear distance $D$. The ground-state energies at the ``chemical'' distances ($D=2/Z$ a.u.) are presented for one-electron dimers with $Z=1\dots100$. Obtained data are compared with the available previous calculations and a good agreement is observed. The comparison of the results for different origin placement of the coordinate system is used as a self-test of the method. The values obtained within the monopole approximation are also presented. It is shown that their dependence on the origin placement can serve to estimate the deviation from the two-center results. 

The developed method, in addition to the energies and wave functions of the ground and lowest excited states, provides the quasicomplete finite spectrum. The Green function computed on the basis of this spectrum gives an access, in particular, to evaluation of the Feynman diagrams within the bound-state QED.

\acknowledgments{Valuable discussions with Ilia Maltsev, Alexey Malyshev, Leonid Skripnikov, and Ilya Tupitsyn are gratefully acknowledged. The work was supported by the Foundation for the Advancement of Theoretical Physics and Mathematics ``BASIS'', by the Russian Foundation for Basic Research (grant number 19-02-00974), by TU Dresden (DAAD Programm Ostpartnerschaften), and by G-RISC.}

\bibliography{main.bib}

\end{document}